\documentclass[sigconf,nonacm]{acmart}

\usepackage{booktabs} 

\usepackage{listings}
\usepackage{xcolor}
\usepackage{makecell}
\usepackage{subfigure}

\definecolor{codegreen}{rgb}{0,0.6,0}
\definecolor{codegray}{rgb}{0.5,0.5,0.5}
\definecolor{codepurple}{rgb}{0.58,0,0.82}
\definecolor{backcolour}{rgb}{0.95,0.95,0.92}

\lstdefinestyle{mystyle}{
    backgroundcolor=\color{backcolour},   
    commentstyle=\color{codegreen},
    keywordstyle=\color{magenta},
    numberstyle=\tiny\color{codegray},
    stringstyle=\color{codepurple},
    basicstyle=\ttfamily\tiny,
    breakatwhitespace=false,         
    breaklines=true,                 
    captionpos=b,                    
    keepspaces=true,                 
    numbers=left,                    
    numbersep=5pt,                  
    showspaces=false,                
    showstringspaces=false,
    showtabs=false,                  
    tabsize=2
}

\usepackage[ruled]{algorithm2e} 

\SetAlFnt{\small}
\SetAlCapFnt{\small}
\SetAlCapNameFnt{\small}
\SetAlCapHSkip{0pt}

\begin{document}

\title[Neural-GASh: A CGA-based neural radiance prediction pipeline for real-time shading]
{Neural-GASh: A CGA-based neural radiance prediction pipeline for real-time shading}


\author{Efstratios Geronikolakis}
\orcid{0000-0002-3974-2816}
\affiliation{%
  \institution{FORTH - ICS, University of Crete, University of Crete}
  \city{}
  \country{}
}
\author{Manos Kamarianakis}
\orcid{0000-0001-6577-0354}
\affiliation{%
  \institution{FORTH - ICS, University of Crete, ORamaVR}
  \city{}
  \country{}
}
\author{Antonis Protopsaltis}
\orcid{0000-0002-5670-1151}
\affiliation{%
  \institution{University of Western Macedonia, ORamaVR}
  \city{}
  \country{}
}
\author{George Papagiannakis}
\orcid{0000-0002-2977-9850}
\affiliation{%
  \institution{FORTH - ICS, University of Crete, ORamaVR}
  \city{}
  \country{}
}
\renewcommand{\shortauthors}{Geronikolakis, Kamarianakis et al.}

\begin{abstract}

This paper presents \textit{Neural-GASh}, a novel real-time shading 
pipeline for 3D meshes, that leverages a neural radiance field architecture 
to perform image-based rendering (IBR) using Conformal Geometric 
Algebra (CGA)-encoded vertex information as input. Unlike traditional Precomputed Radiance Transfer (PRT) methods, that require expensive offline precomputations, our learned model directly consumes CGA-based representations of vertex positions and normals, enabling dynamic scene shading without precomputation. 
Integrated seamlessly into the Unity engine, Neural-GASh facilitates accurate shading of animated and deformed 3D meshes—capabilities essential for dynamic, interactive environments. The shading of the scene is implemented within Unity, where rotation of scene lights in terms of Spherical Harmonics is also performed optimally using CGA. 
This neural field approach is designed to deliver fast and efficient light transport 
simulation across diverse platforms, including mobile and VR, while preserving 
high rendering quality. Additionally, we evaluate our method on scenes generated 
via 3D Gaussian splats, further demonstrating the flexibility and robustness of 
Neural-GASh in diverse scenarios. Performance is evaluated in comparison to 
conventional PRT, demonstrating competitive rendering speeds even with complex geometries.

\end{abstract}

\maketitle

\section{Introduction}
\begin{figure*}[tbp]
  \centering
  \includegraphics[width=1\textwidth]{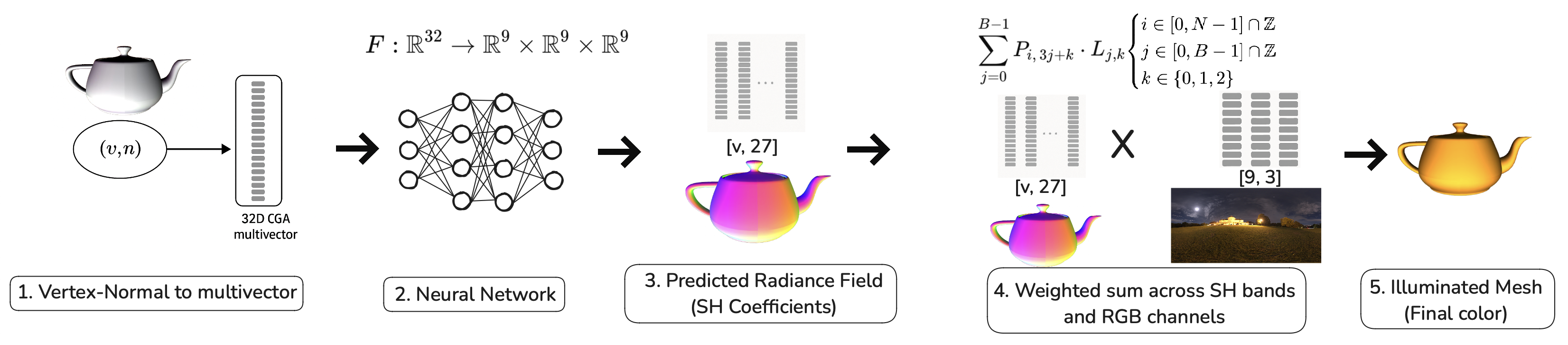}
  \caption{The neural radiance field architecture utilized in our pipeline.}
  \label{fig:architecture}
\end{figure*}

Global illumination describes a comprehensive shading technique that 
realistically simulates both lighting and reflections by accurately modeling the
behavior of light in a scene \cite{whitted2020origins}. It is widely utilized
across Computer Graphics (CG) applications to enhance the realism of rendered scenes 
with the resulting visual complexity largely determined by the specific light transport algorithm employed. Rendering realistic lighting effects in real-time interactive CG applications remains challenging and computationally demanding task, particularly when aiming for highly accurate global illumination. Accurate capture of the complex light interactions within a scene --such as indirect illumination, reflections, and soft shadows --typically demands significant computational 
resources. PRT has been developed as an effective method to manage these complexities by precomputing the interactions of light with static scenes under varied lighting conditions, facilitating interactive rendering of sophisticated global illumination effects. However, traditional PRT approaches \cite{Sloan2002}
encounter significant limitations, primarily due to their reliance on extensive
offline computations. The time required for these precomputations varies
significantly, depending on the geometric complexity and the computational
capabilities of the hardware utilized. This limitation intensifies considerably
in dynamic scenes where geometry is updated at runtime. In that respect, scenarios
involving animations, geometric deformations, or changes in material properties
require repeated precomputations, making the pipeline impractical, especially on
lower-performance hardware. Consequently, conventional PRT algorithms are fundamentally limited due to the requirement for static scene geometry.
By contrast, the Blinn-Phong shading model \cite{blinn1977models} is a widely
adopted, computationally inexpensive approach used in many real-time
applications for approximating local illumination. However, by considering only direct lighting, the model fails to account global illumination effects such as
soft shadows, indirect light bounce, and color bleeding, limiting its ability to
produce photorealistic results.

\begin{figure}[tbp]
    \centering
    \includegraphics[width=1\linewidth]{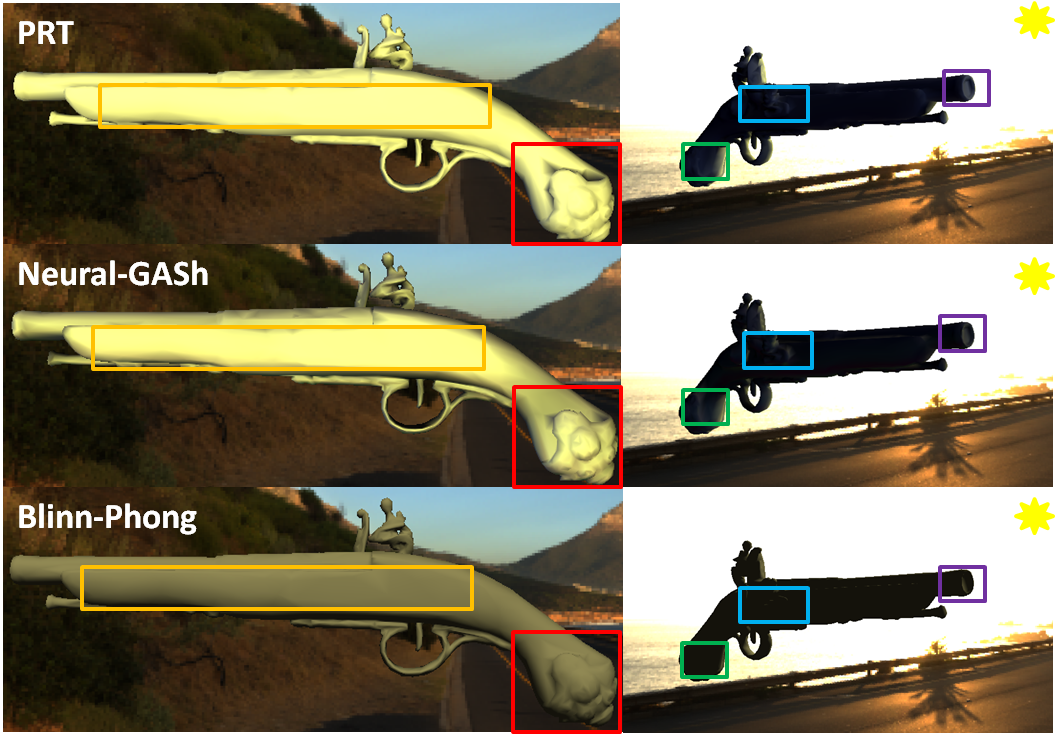}
    \caption{Comparison of pistol model rendered using traditional PRT, Neural-GASh, and Blinn-Phong algorithms. Neural-GASh closely matches PRT in crevices (red), exposed areas (yellow), and low-light regions (green, blue, purple). The sun icon indicates the main light source direction. Blinn-Phong shows uniform lighting from a manually added source, resulting in an entirely dark back side due to the absence of ambient illumination.}
    \label{fig:pistola}
\end{figure}

To address these limitations, we propose \textit{Neural-GASh}, 
a \textit{neural field} \cite{10.1111:cgf.14505} 
(see Section~\ref{sec:our_method} and Figure~\ref{fig:architecture} ) 
approach that replaces the rigid precomputation stage of 
traditional PRT with a flexible neural network capable of predicting 
radiance transfer in real time. Rather than relying on heavy offline 
sampling and Spherical Harmonic (SH) coefficient generation for every 
static or animated mesh, our method learns a direct mapping from mesh 
features -- specifically CGA-based representations of vertex positions 
and normals -- to their corresponding 
SH lighting coefficients. In that respect, Neural-GASh eliminates 
the need for repeated precomputation of PRT coefficients when the geometry 
changes, offering a lightweight and scalable solution suitable for dynamic 
and deformable scenes. Although the approach does not fully match the photorealism and 
precision of traditional PRT, it significantly improves upon local 
illumination models such as Blinn-Phong (see Figure~\ref{fig:pistola}). Despite being widely 
adopted due to computational efficiency, such models fail to account for global illumination 
effects. Notably, Neural-GASh supports real-time rendering 
of animated rigged models, with CGA facilitating efficient 
SH operations for dynamic lighting control, particularly valuable for 
interactive CG applications and performance\--constrained environments.

Specifically, Neural-GASh offers the following key capabilities:

\vspace{-4mm}

\begin{enumerate}
    \item \textbf{Geometry-Agnostic Operation:} Neural-GASh is capable of operating on 
    any type of 3D geometry, whether simple or highly complex. Leveraging the 
    representational power of neural networks, it can predict per-vertex SH 
    coefficients in a fraction of the time required by traditional methods.
    
    \item \textbf{Trained using CGA-Based data:} The training data for 
    Neural-GASh training data comprises CGA-based multivectors encoding vertex coordinates and normals.
    This richer geometric 
    representation enhances the model’s ability to learn lighting behavior 
    across complex spatial configurations, improving (visual) accuracy 
    over traditional vector-based inputs (see Figure~\ref{fig:cats}).

    \item \textbf{Cross-Platform Compatibility:} While developed and trained  
    in Python, Neural-GASh maintaints full compatibility with Unity game engine. This facilitates seamless integration into interactive applications 
    and enables deployment across a wide range of platforms, including 
    resource constrained mobile devices. By eliminating the costly PRT 
    precomputation step, Neural-GASh significantly reduces the hardware requirements for real-time rendering.

    \item \textbf{Dynamic Environmental Lighting:} The system supports realistic 
    environmental lighting manipulation by efficiently rotating the spherical harmonic coefficients of the radiance map. This is accomplished through CGA, which enables smooth and computationally inexpensive lighting transformations.

    \item \textbf{Support for Deformable and Animated Geometry:} Neural-GASh adapts to runtime geometry changes by efficiently recomputing per-vertex 
    coefficients when deformations or animations occur. This enables robust handling 
    of dynamic scenes with continuously changing vertex positions and normals.

    \item \textbf{Real-Time VR Performance:} Neural-GASh achieves real-time rendering performance in Virtual Reality (VR) environments for typical interactive scenes. This is particularly effective when applied to lightweight 
    3D meshes containing fewer than 1000 vertices, enabling smooth illumination 
    updates even on performance-constrained VR hardware.
\end{enumerate}

\section{Related Work}
\subsection{Precomputed Radiance Transfer (PRT)}

PRT \cite{Sloan2002} is a widely adopted technique in global illumination (GI)
that enables real-time rendering of complex lighting phenomena by precomputing
how light interacts with static geometry. It efficiently captures soft shadows,
interreflections, and other low-frequency lighting effects using SH,
significantly reducing runtime costs. However, PRT suffers from several critical
limitations that restrict its applicability in dynamic or real-time
environments. Most notably, the geometry of the scene must remain static; any
changes --such as deformations, animations, or topology updates -- require
repeating the entire precomputation process. This offline precomputation step is
often time-consuming and scales poorly with geometric complexity. Furthermore,
PRT does not adapt efficiently to changes in material properties or lighting
environments at runtime, limiting its flexibility in interactive applications.
These constraints pose significant challenges for integrating PRT into modern
rendering pipelines, especially on resource-constrained platforms like mobile or
embedded devices.

\subsection{Approaches to Precomputed Radiance Transfer}

In the ongoing pursuit of more efficient and mathematically elegant global
illumination techniques, an advanced formulation of the PRT algorithm has
emerged \cite{papaefthymiou2018real} through its integration with CGA \cite{hitzer2012introduction,papaefthymiou2016inclusive,sommer2013geometric}. 
This approach leverages the representational power
of CGA to optimize the computation of radiance transfer, enabling real-time
rendering with reduced overhead while preserving visual fidelity. At its core,
CGA-based PRT represents objects and transformations using multivectors—uniform
algebraic entities capable of encoding geometric transformations in a compact
and consistent framework. This formulation allows for the rotation of SH
coefficients using CGA rotors, which consist of only four parameters. However,
despite these advantages, the CGA-enhanced PRT model inherits several of the
core limitations of traditional PRT. Most notably, it lacks support for
translation and scaling of SH coefficients, thereby limiting its use in dynamic
scenes involving object movement or geometry deformations. Additionally, this
model remains constrained to static geometries and does not natively accommodate
animations or vertex-level transformations. While CGA-based PRT represents a
notable advancement in rendering static environments with complex lighting
interactions, its applicability remains limited in real-time scenarios that
demand geometric flexibility. 

PRT textures \cite{dhawal2022prtt} extend the traditional PRT framework by
storing precomputed transfer values directly in texture maps, enabling
high-quality rendering of both diffuse and glossy reflections—even on
low-tessellation meshes. This approach achieves a compelling balance between
visual fidelity and performance, supporting real-time rendering of complex
lighting effects. However, since interreflections and material responses are
baked into the textures, dynamic changes to geometry or lighting require costly
precomputation, limiting the method's flexibility in interactive scenarios.

Another method for real-time global illumination using PRT and spherical
harmonics to approximate complex lighting effects like indirect illumination and
soft shadows is the MoMo-PRT \cite{schneider2017efficient}. This method achieves
interactive performance through carefully optimized data structures and avoids
the use of deep learning or neural networks, focusing instead on classical
computer graphics techniques. However, the method is limited by its reliance on
precomputation, which reduces flexibility for highly dynamic scenes, and may
struggle with high-frequency lighting or geometry, while memory usage can remain
a challenge in complex environments.

An approach proposed in \cite{thul2020precomputed} enables accurate scene decomposition into surface
reflectance and lighting, supporting advanced effects like occlusion and
indirect illumination. By leveraging precomputed data from geometry and images,
the method enhances realism in relighting and content creation. The method improves
rendering quality by incorporating indirect light but may face limitations due
to spherical harmonics, which can struggle with highly complex scenes.

\subsection{Neural Rendering}
In \cite{li2019deep} DeepPRT, a deep convolutional neural network designed to
efficiently encapsulate the radiance transfer characteristics of freely
deformable objects for real-time rendering is introduced. Traditional PRT
methods often struggle with dynamic deformations due to their reliance on static
precomputations. DeepPRT addresses this limitation by learning a compact
representation that can adapt to various deformations, enabling real-time
rendering of complex lighting effects such as soft shadows and interreflections
on deformable objects. This approach leverages the power of deep learning
networks to generalize across different deformation states, offering a more
flexible and efficient solution for rendering deformable objects under dynamic
lighting conditions. However, the deep learning network used in this approach
appears to have potential for further optimization in terms of memory usage and
computational performance.

The method proposed in \cite{rainer2022neural} integrates neural networks with traditional PRT to
achieve real-time global illumination in static scenes with dynamic lighting.
Their neural architectures, inspired by PRT, learn light transport efficiently
without requiring ray-tracing hardware. Results show improved illumination
predictions with real-time performance. However, the method operates on images,
not 3D geometry, learning a mapping from lighting to rendered images, which
limits its ability to handle dynamic geometry. This image-based approach trades
geometric flexibility for efficiency, making it less suitable for real-time
object manipulation or scene changes.

\subsection{Gaussian Splatting}

3D Gaussian Splatting (3DGS) has emerged as a significant advancement in
computer graphics, offering an explicit scene representation that facilitates
real-time rendering and efficient scene manipulation \cite{fei20243d}. Unlike
traditional implicit neural radiance field approaches, 3DGS utilizes millions of
learnable 3D Gaussians to model scenes, enabling high-quality reconstructions
with computational efficiency. \cite{kerbl20233d} introduced a real-time 3D
Gaussian Splatting pipeline that efficiently reconstructs high-quality scenes
from multi-view images, demonstrating its superiority over traditional neural
rendering methods in terms of both speed and quality.

Recent studies have explored various aspects of 3DGS. For instance,
\cite{chen2025survey3dgaussiansplatting} provides a comprehensive review of
them. The paper explores the foundations, recent advances, and practical
applications of 3DGS, particularly in areas like virtual reality and interactive
media. It also offers a comparative evaluation of leading models and outlines
key challenges and future research directions in this emerging field.

Finally, \cite{zhang2024prtgaussian} presents a
real-time, relightable novel-view synthesis method that combines 3DGS with PRT,
leveraging high-order spherical harmonics for efficient and detailed
relighting. It reconstructs coarse geometry from multi-view images and refines
both geometry and light transport per Gaussian for high-quality results.
However, the method is limited to diffuse materials, lacks specular and
view-dependent effects, and struggles with high-frequency details like sharp
shadows due to the low-frequency nature of spherical harmonics.

\section{Our Method: Neural-GASh }
\label{sec:our_method}
Neural-GASh is a novel framework grounded in the core principles of PRT, designed to
extend the traditional algorithm’s capabilities and broaden its applicability.
Its primary objective is to modernize and enhance the PRT pipeline by
introducing support for dynamic features such as geometry animations,
deformations -- capabilities traditionally unsupported
due to the static nature of precomputed data. Additionally, Neural-GASh addresses the
computational limitations of conventional PRT by significantly reducing the
overhead of the precomputation phase, thereby enabling deployment across a wide
range of devices, including resource-constrained platforms.

The concept of PRT has been extensively explored within the research community
and applied in various contexts. However, to the best of our knowledge, there
has not yet been a comprehensive and deployable PRT pipeline that fully
leverages neural networks for integration into real-time 3D applications and
games, particularly across a wide range of hardware platforms, including both
desktop and mobile devices. Neural-GASh addresses this gap by providing a lightweight,
efficient, and platform-agnostic solution that is fully compatible with
Unity—one of the most widely adopted game engines in industry and academia.
The plug-and-play design facilitates seamless integration into existing
rendering pipelines, enabling realistic lighting effects that adapt dynamically
to the environment and significantly enhance the visual fidelity of 3D geometry.

\begin{figure}[tbp]
    \centering
    \includegraphics[width=1.0\linewidth]{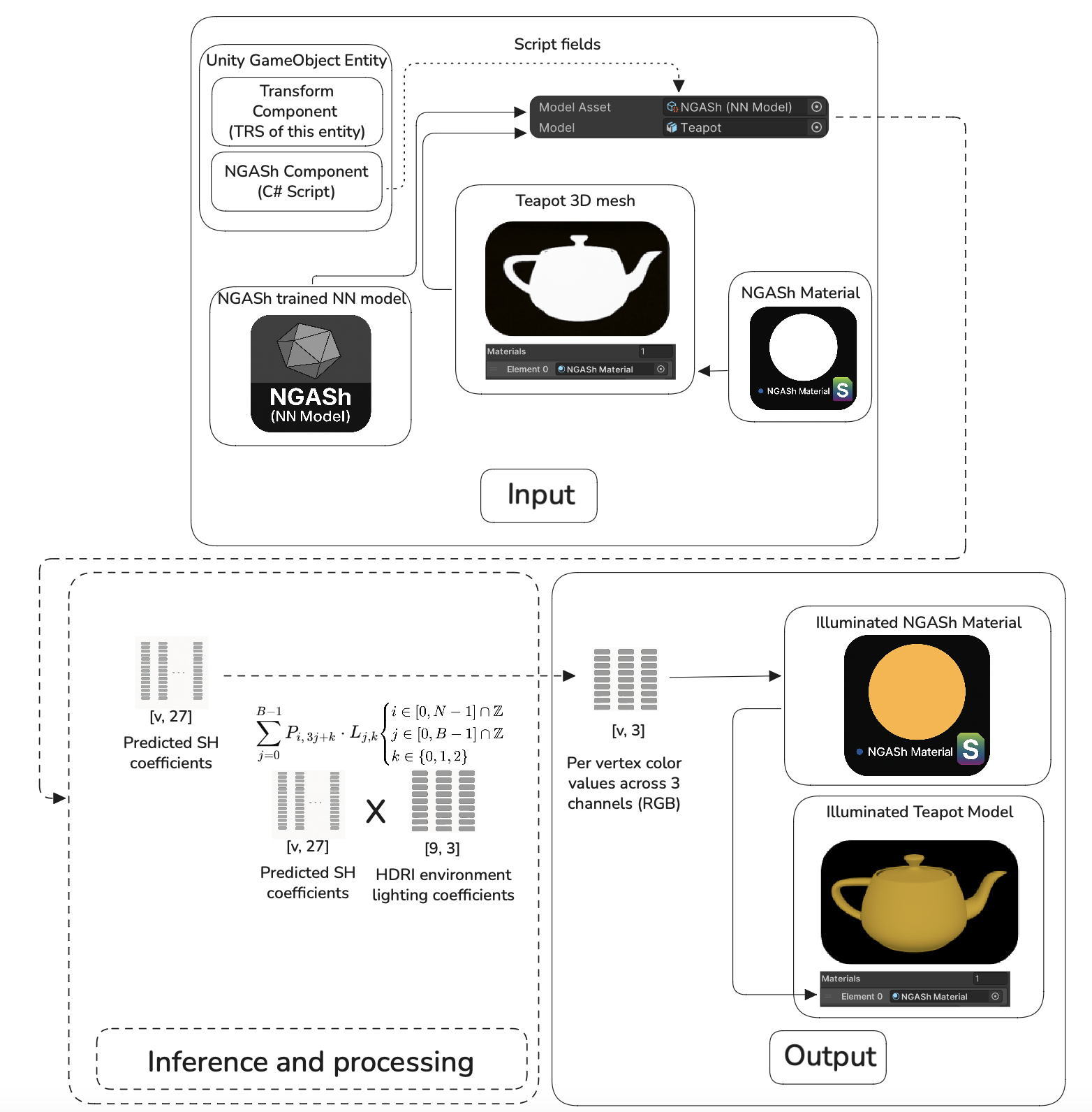}
    \caption{The main structure of the Neural-GASh pipeline in Unity for the teapot model.}
    \label{fig:unity_pipeline}
\end{figure}

Figure~\ref{fig:unity_pipeline} depicts the rendering pipeline of the
Neural-GASh (NGASh) system as deployed within the Unity environment.
At runtime, the pipeline is initialized via a custom Unity component
that coordinates the model inference and shading process. Two core
inputs are required: a 3D mesh (e.g., a teapot) and a neural model
trained offline in Python to predict spherical harmonic (SH) radiance
coefficients from Conformal Geometric Algebra (CGA) representations.

For each vertex in the mesh, the position and normal are encoded as a
32-dimensional multivector using CGA. These multivectors are passed to
the NGASh neural field, which outputs 27 SH coefficients per
vertex—capturing the radiance response across three color channels and
multiple frequency bands. The predicted coefficients are then
projected onto lighting basis functions derived from a
high-dynamic-range (HDRI) environment map via a bandwise inner
product, as formalized in Equation~\ref{eq:SHCoeff}.

The resulting per-vertex RGB values are transferred to the GPU and
used to update the mesh appearance in real time through a lightweight
fragment shader. This approach enables dynamic, data-driven lighting
effects directly on mesh geometry, demonstrating the practical
integration of neural light transport fields and CGA-based geometric
encoding within interactive rendering environments.

Our architecture adheres to the principles of neural fields as
introduced in recent literature \cite{10.1111:cgf.14505}.
Specifically, we model a continuous function \( f: \mathbb{R}^{32}
\rightarrow \mathbb{R}^{9} \times \mathbb{R}^{9} \times \mathbb{R}^{9}
\), where the input is a 32-dimensional CGA multivector that encodes
the local geometric context of a vertex—including position and
normal—and the output is a set of SH coefficients: 9 coefficients per
color channel, corresponding to 3 SH bands. This formulation enables
the network to learn a continuous radiance field defined over the mesh
surface, consistent with the neural field paradigm, where spatially
varying scene properties are regressed directly from embedded
geometric descriptors.

To compute the shaded color at each vertex, the predicted SH coefficients 
\( P_{i,\,3j + k} \) are projected onto the SH lighting basis \( L_{j,k} \) 
using a per-channel inner product:
\begin{equation}
C_{i,k} = \frac{1}{255} \sum_{j=0}^{B-1} P_{i,\,3j + k} \cdot L_{j,k},
\label{eq:SHCoeff}
\end{equation}
\noindent where \( i \in [0, N\!-\!1] \cap \mathbb{Z} \), 
\text{for N number of vertices} \( j \in [0, B\!-\!1] \cap \mathbb{Z} \), 
\text{for $B=b^2$ where $b$ is the SH bands} and \( k \in \{0, 1, 2\} \), 
for the 3 color channels (RGB).

This projection yields a fully shaded mesh under environment lighting. 
As such, Neural-GASh operates as a neural light transport field, 
learning radiance responses directly from CGA-encoded geometry.

Neural-GASh is designed to operate across all types of 3D meshes --
whether rigid or deformable, and regardless of geometric complexity. 
Given the mesh's vertex
positions and normals, the system rapidly approximates the per-vertex radiance
transfer coefficients that would traditionally be computed through an expensive
offline PRT precomputation step. Once these coefficients are generated, they are
combined with the scene's lighting coefficients to determine the final color at
each fragment. This color is then assigned to the corresponding vertex, enabling
the rendering of realistic lighting effects. The model runs once at
initialization and is only re-evaluated when a change in the geometry is
detected. Since the per-vertex SH coefficients remain constant unless the vertex
positions or normals are modified—such as during animation or deformation—the
runtime computational cost is minimized, enabling efficient performance even in
dynamic scenes.

To simplify deployment and integration, Neural-GASh operates directly on 3D mesh
geometry without requiring UV maps or geometry images. It is compatible with
both textured and untextured assets and supports 
real-time texture updates, with
lighting dynamically adapting to the scene. For rotating the SH lighting
environment, the original design utilizes CGA to
perform spherical harmonic coefficients rotation through rotor-based
transformations. However, since Unity does not natively support CGA libraries
or math constructs, we adapted our implementation to use quaternions as an
equivalent representation. While mathematically distinct, quaternion-based
rotation preserves the behavior and correctness of the lighting manipulation.
This Unity-based implementation opens the door to a wide range of
applications—including augmented and mixed reality—where real-time lighting can
greatly enhance immersion and realism.

\section{Implementation Details \& Datasets}
Neural-GASh employs a fully connected feedforward neural network tailored for per-vertex
regression from geometric descriptors to spherical harmonic (SH) lighting
coefficients. Each vertex-normal pair of the mesh is encoded as a 32-dimensional feature vector,
corresponding to a CGA multivector. These
multivectors are constructed by transforming each vertex-normal pair using the
geometric algebra (GA) framework provided by the Elements project
\cite{Elements2023}, implemented in Python. This encoding captures both position
and orientation information in a unified algebraic form, enabling the network to
learn a mapping from geometric structure to lighting representation.

To represent the geometric relationship between a vertex and its associated
normal vector in the framework of CGA, we construct a \emph{motor}, i.e., a versor
that encodes both rotation and translation, based on the vertex-normal pair
$(\mathbf{v}, \mathbf{n}) \in \mathbb{R}^3 \times \mathbb{R}^3$. 
The normal
vector $\mathbf{n}$ is first normalized to unit length, defaulting to the upward
direction $(0,1,0)$ in degenerate cases. 
A CGA rotor $R$ is constructed from the unit quaternion $q$ that aligns $\mathbf{y} = (0,1,0)$ 
with the normal vector $\mathbf{n}$. A CGA translator $Tr$ corresponding to the translation by $v$ 
is then multiplied with $R$, resulting in a 32-dimensional \textit{motor} $M$ which 
captures the full rigid body transformation implied by the vertex-normal pair.
This representation provides a geometrically meaningful descriptor suitable for
learning-based algorithms and geometric processing tasks.

From motor $M$ we can still extract 
a corresponding transformation matrix $T$ that is used by Unity shaders. The 
relationship between $T$ and  $M$ is well established \cite{dorst2009geometric,kamarianakis2021}, 
with conversion procedures available in the GA-Unity package \cite{kamarianakis2024ga} 
used within the Unity environment.

Following this transformation, each vertex-normal pair of the mesh is represented as a
32-dimensional multivector in the CGA framework. These multivector
representations serve as inputs to a neural network, which learns a mapping from
the 32-dimensional latent space to a 27-dimensional output space corresponding
to spherical harmonics (SH) coefficients, as commonly used in radiance transfer
approximation~\cite{green2003spherical}. To ensure stable optimization and
improve generalization, batch normalization is applied to the input and
intermediate layers. Additionally, dropout regularization is incorporated with a
decaying schedule across network layers to mitigate overfitting while preserving
representational capacity.

\begin{figure}[tbp]
    \centering
    \subfigure[Neural-GASh: 32D CGA input ]{\includegraphics[trim=0 30 0 20, clip, width=0.235\textwidth]{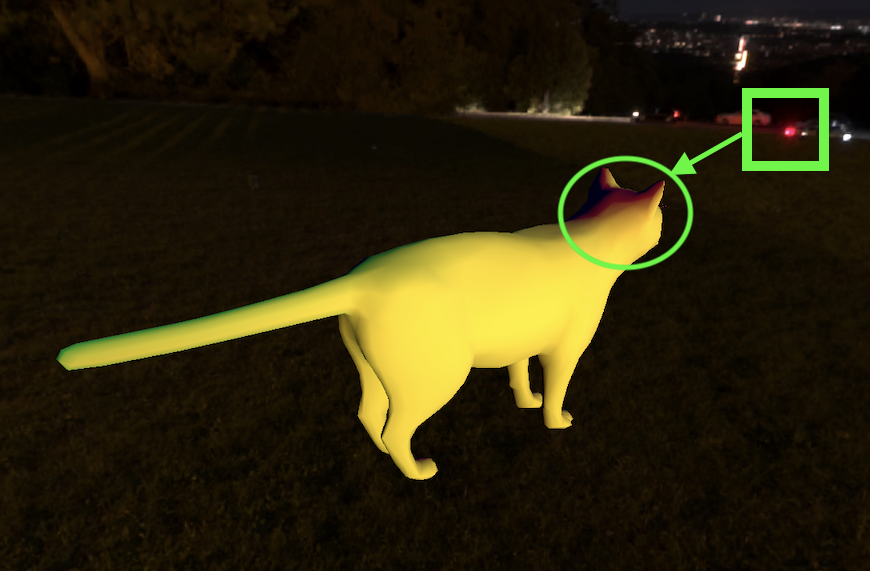}}
    \hfill
    \subfigure[Same architecture network: 6D euclidean input]{\includegraphics[trim=0 30 0 14, clip, width=0.235\textwidth]{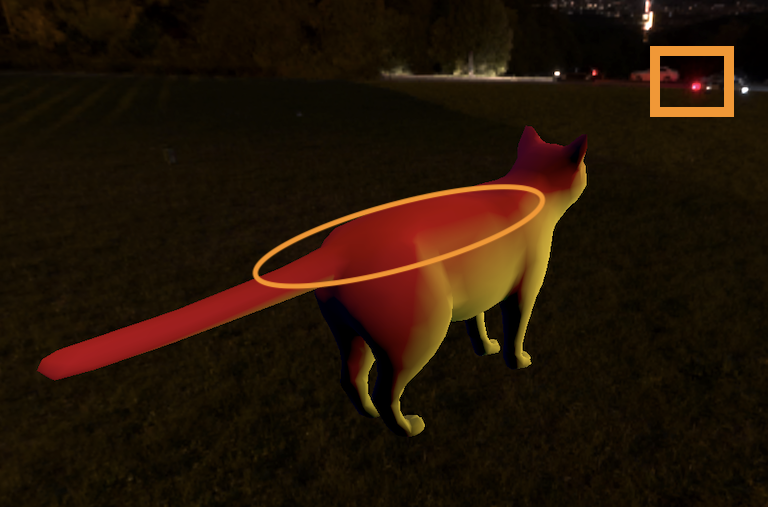}}
    \caption{Shading results of a cat mesh using (a) Neural-GASh, i.e., model is  
    using 32 CGA multivectors as input, and (b) a network 
    with identical architecture as Neural-GASh, using 6D euclidean coordinates as input (for position and normal of each vertex). In (a), the red light source illuminates only the edge of the cat's head, consistent with its frontal position. In contrast, in (b), the entire back of the cat (including the tail) is unrealistically illuminated by the same frontal red light source. This represents a challenging edge case in large-scale 3D meshes, which is more accurately addressed by incorporating the 32D Conformal Geometric Algebra (CGA) representation in our neural network input.}
    \label{fig:cats}
\end{figure}

We opted to train our neural network using 32-dimensional multivector
representations derived from CGA, as this approach yielded more stable and
perceptually accurate results, particularly in challenging scenarios. 
Using a 32-dimensional multivector as input provides a richer and more expressive 
representation of local geometry compared to a simple 6D vertex-normal pair. 
Multivectors can encode higher-order spatial relationships and directional context, 
allowing the network to better capture subtle variations in surface structure 
and lighting interactions, leading to improved prediction accuracy.
This representation offers inherent advantages, including equivariance to rigid 
body transformations and
the ability to capture nonlinear spatial relationships pertinent to radiative
transport. As can be observed in Figure~\ref{fig:cats}, networks trained on 
6D euclidean-based inputs
produced suboptimal results in cases involving large-scale or complex geometry,
whereas the CGA-based formulation demonstrated superior visual performance and
generalization in such conditions.

\begin{figure*}[tbp]
    \centering
    \includegraphics[width=1\linewidth]{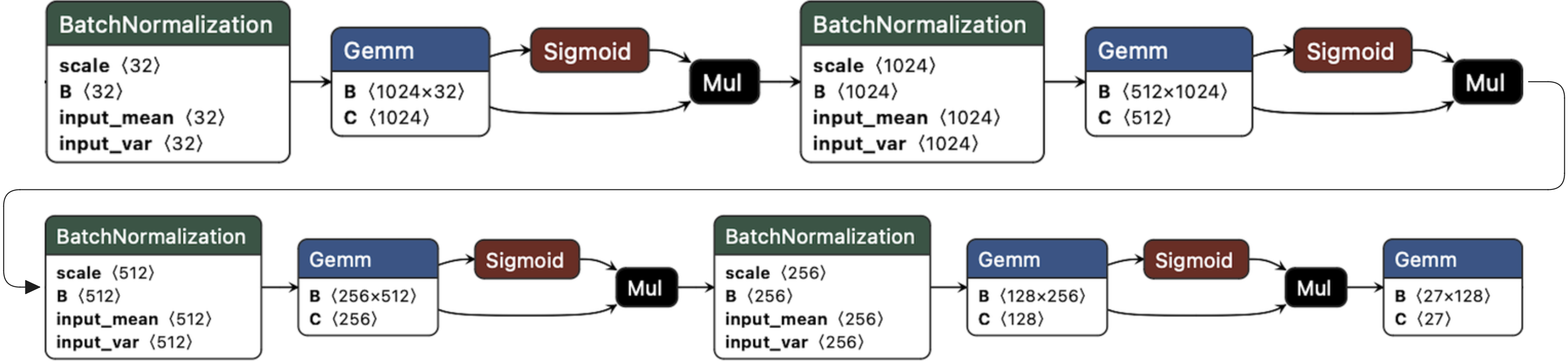}
    \caption{\textbf{The neural network architecture, used in the neural radiance field of Neural-GASh.}}
    \label{fig:nn}
\end{figure*}

The architecture, as shown in Figure~\ref{fig:nn}, consists of five fully 
connected layers with decreasing dimensionality: 1024, 512, 256, 128, and 
finally 27 output neurons. Each hidden layer is followed by a SiLU
(Sigmoid-weighted Linear Unit) activation function, selected for its smooth
non-linearity and suitability for regression tasks. The final 27-dimensional
output is reshaped to match the number of vertices, allowing the model to
generalize across arbitrary mesh sizes. This architecture strikes a balance
between expressive capacity and computational efficiency, enabling deployment in
real-time and on resource-constrained platforms.

\begin{figure}[tbp]
    \centering
    \subfigure[Unshadowed version]{\includegraphics[trim=0 50 0 50, clip, width=0.235\textwidth]{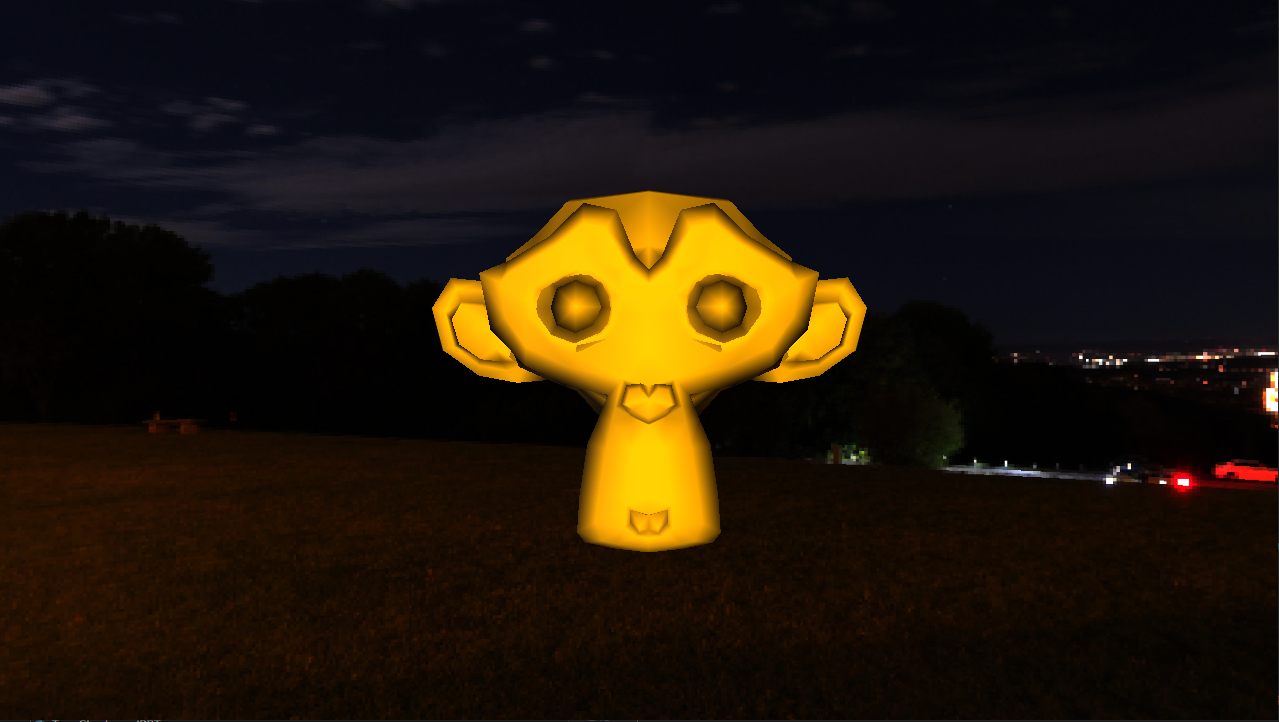}}
    \hfill
    \subfigure[Shadowed version]{\includegraphics[trim=0 50 0 50, clip, width=0.235\textwidth]{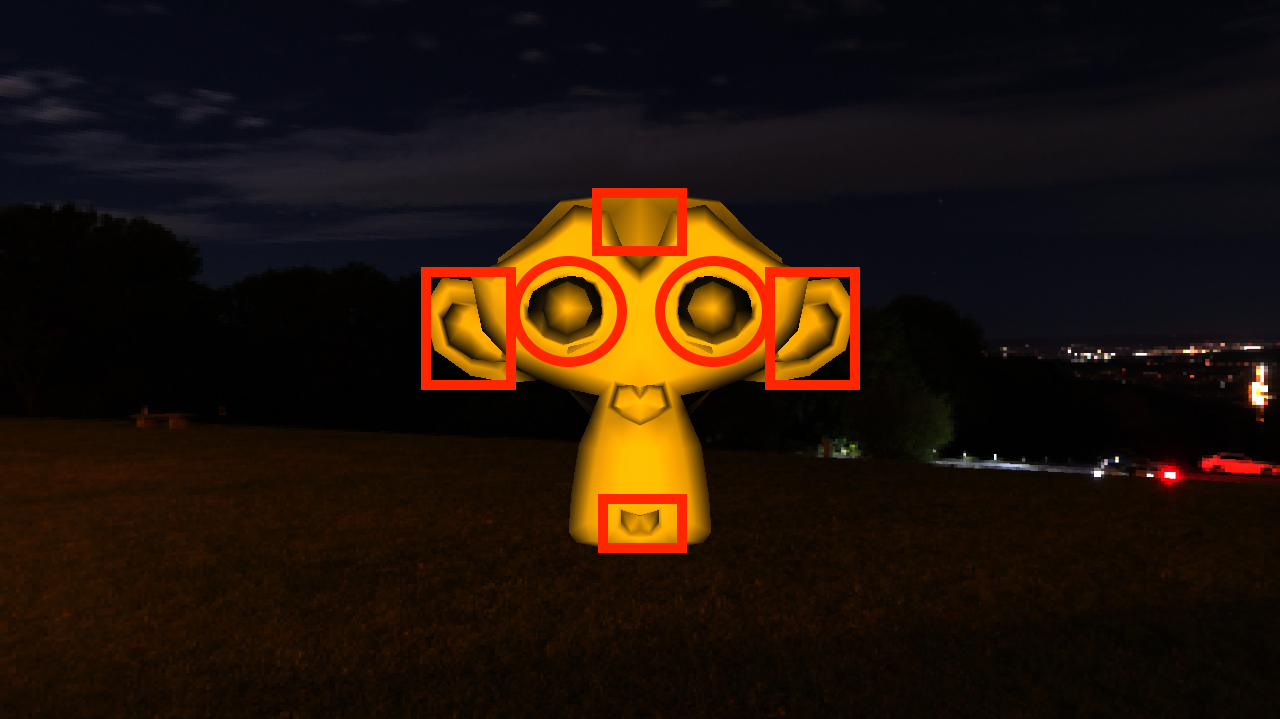}}
    \caption{Traditional PRT results on the monkey mesh: (a) unshadowed and (b) shadowed. In (b), regions with increased shadowing (primarily crevices) are highlighted with red squares. The overall darkening of the surface in the shadowed version reflects visibility-based occlusion, indicating correct shadow integration.}
    \label{fig:monkeys}
\end{figure}

Training data was generated using a custom Python implementation of the
traditional PRT algorithm. As no publicly available Python-based PRT
implementation exists to our knowledge, we reimplemented both the unshadowed and
shadowed variants from scratch. An example can be viewed in Figure~\ref{fig:monkeys}. 
Our method was built as an extension to the
PyGandalf framework \cite{petropoulos2024pygandalf}, leveraging its rendering
and mesh processing capabilities. The implementation follows the principles of
the original C++ method described in \cite{c-he2018PRT-SH}, adapted for modern
Python-based neural network workflows.

Using this pipeline, we processed a collection of 3D meshes with
varying geometric complexity. Each mesh contained vertex positions and
normals; if normals were missing, they were computed programmatically.
For each mesh, the ground-truth spherical harmonic (SH) coefficients
were computed using the traditional PRT method and exported to a text
file for supervised training. Each file contained \(27 \times N\)
floating-point values, where \(N\) is the number of vertices. The
$i$-th row corresponds to the $i$-th vertex and stores 27 SH
coefficients, denoted \(t_0\) to \(t_{26}\), representing the
vertices' radiance response to incident lighting. These coefficients
provide a compact, rotation-aware encoding of light transport and
remain consistent across different meshes, provided a fixed set of
lighting samples and SH bands.

We employed a hemispherical sampling grid with 5 samples per axis (25
lighting directions total) and projected lighting onto 3 SH bands, resulting in
27 SH coefficients per vertex across RGB channels. This setup balances visual
fidelity and computational efficiency. Accordingly, Neural-GASh is trained to
approximate traditional PRT coefficients under this configuration. 

The dataset used to train Neural-GASh is available at \\
\href{https://github.com/stratosger/NeuralGASh-Dataset}{https://github.com/stratosger/NeuralGASh-Dataset}.

Once the 27 SH coefficients are predicted per vertex, we compute a weighted sum
across SH bands and RGB channels by multiplying them with corresponding lighting
coefficients derived from the HDRI environment map. While conceptually similar
to a dot product, this operation accounts for the structured nature of SH
coefficients and their alignment across RGB channels. These lighting
coefficients are critical, as they determine the diffuse contribution of scene
illumination.

Each HDRI environment map is processed using a custom Python based PRT implementation to
compute and store the corresponding SH coefficients in separate files, enabling seamless
runtime integration within Unity. The light projection function computes the
spherical harmonic (SH) coefficients \cite{ramamoorthi2001efficient}
representing an environment light source by projecting the incoming radiance \(
L(\omega) \) onto the SH basis. Given a set of \( N \) uniformly distributed
sample directions \( \{\omega_i\}_{i=1}^N \), the SH coefficient \( \mathbf{c}_l
\in \mathbb{R}^3 \) for each basis function \( Y_l(\omega) \), where \( l = 0,
\dots, B^2 - 1 \) and \( B \) is the number of SH bands, is approximated as: 
$\displaystyle\mathbf{c}_l \approx \frac{4\pi}{N} \sum_{i=1}^{N} L(\omega_i) \cdot
Y_l(\omega_i),$    
where \( L(\omega_i) \) is the RGB radiance obtained from a
light probe in direction \( \omega_i \). The scaling factor \( \frac{4\pi}{N} \) ensures correct
integration over the unit sphere under uniform sampling.

After training process is completed, the model is exported to the ONNX format,
compatible with Unity’s Barracuda package -- Unity’s built-in runtime for neural
network inference. A key challenge during model integration involved differences
in input data layout: while PyTorch uses the [batch, vertices, features] format,
Barracuda expects [batch, height, width, channels]. This required reordering
input tensors to ensure correct output. Proper alignment between SH coefficients
and vertex indices is critical, as it directly impacts the accuracy of
per-vertex lighting. Furthermore, due to the inability to execute native Python
code within the Unity environment, we employed the GA-Unity package
\cite{kamarianakis2024ga} to perform the transformation of vertex-normal pairs
into multivector representations.

To overcome Barracuda’s 2GB tensor size limit and support high-resolution
meshes, we instantiated separate Neural-GASh model instances, one for each mesh. These
instances run simultaneously, allowing the system to handle complex or large-scale
geometries without exceeding memory constraints. Moreover, to optimize real-time
rendering of Neural-GASh in Unity, we implemented several targeted strategies to
improve both computational efficiency and visual responsiveness. First, to avoid
redundant computation, we cache vertex normals and lighting coefficients between
frames and only re-run the neural network model if either geometry or lighting
has changed. Moreover, we further reduce computational overhead by detecting
rapid model motion via a velocity threshold; when the object moves quickly, the
model update is temporarily skipped under the assumption that the lighting
contribution will be perceptually negligible. To leverage hardware acceleration,
the expensive color reconstruction step—where spherical harmonic coefficients
are combined with lighting—is offloaded to the GPU using a compute shader. This
parallelizes per-vertex operations and significantly reduces CPU load.
Additionally, the geometric encoding stage, where vertex-normal pairs are
transformed into 32-dimensional CGA multivectors, is
parallelized using Unity's job system and NativeArray types to ensure efficient
multithreaded execution.

\section{Results \& Discussion}

Neural-GASh was evaluated using a consumer-grade workstation, representative of typical 
mid-range gaming or development setups. \textbf{Processor:} AMD Ryzen 5 2400G (Raven Ridge, 14nm Technology),
\textbf{Memory:} 8 GB Dual-Channel RAM @ 2666 MHz,
\textbf{Graphics Processing Unit (GPU):} NVIDIA GeForce GTX 1060 (6GB VRAM).

\subsection{Traditional PRT vs Neural-GASh}

This benchmark compares the computational efficiency of the two approaches
(traditional precomputed radiance transfer (PRT) and our Neural-GASh method) under
identical hardware conditions, providing insight into their relative
performance.

Since the traditional PRT algorithm was implemen\-ted in Python, we utilized the
time.time() function as documented in the official Python Standard Library. This
function accurately measures code execution time, making it suitable for
benchmarking purposes.

For Neural-GASh, execution time was measured with\-in the Unity game engine, which
serves as the intended deployment environment for the neural model. Integration into Unity demonstrates the method's practical applicability in real-time contexts,
including interactive rendering and XR systems. To achieve high-precision timing,
we employed the Stopwatch class from the .NET framework. As a high-resolution
timer specifically designed for benchmarking and performance analysis, Stopwatch
provides precise measurements of elapsed time and is well-suited for profiling
code execution within Unity.

\begin{figure}[tbp]
    \centering
        \includegraphics[width=1\linewidth]{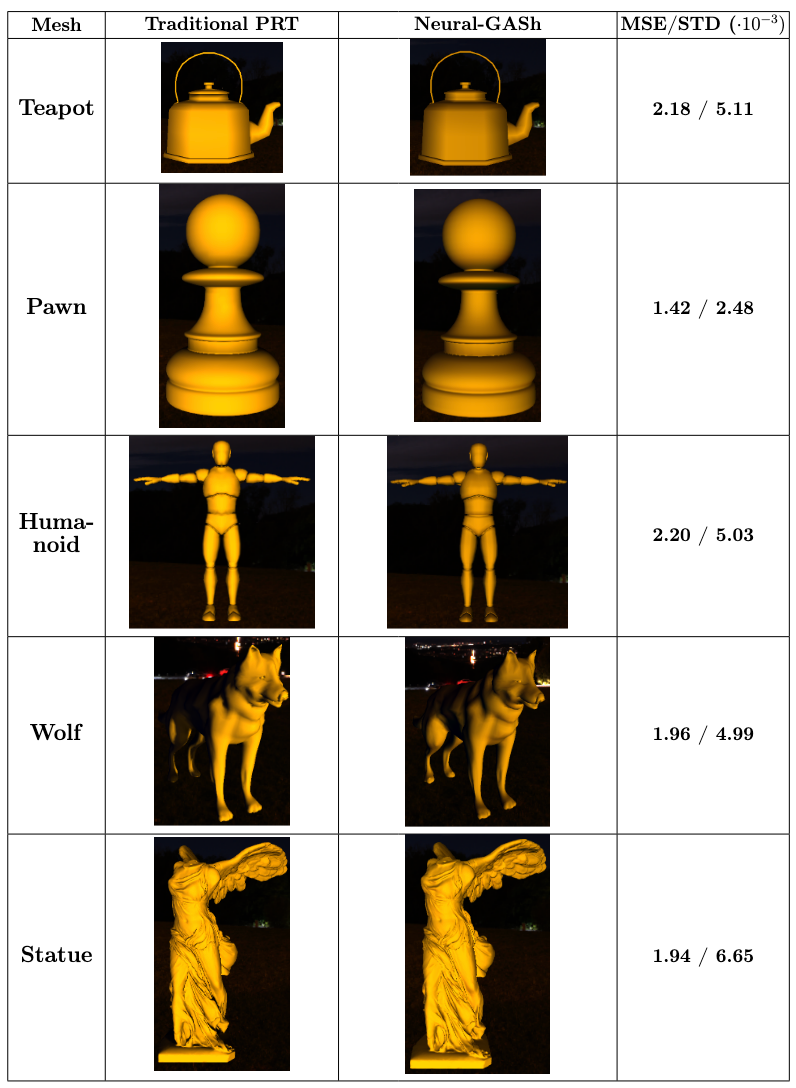}
    \caption{Comparison between traditional PRT and Neural-GASh results. The close visual match demonstrates the effectiveness of our approach. Minor darkening in the Neural-GASh output, due to coefficient normalization in the Python pipeline, is corrected via an intensity multiplier in the shader. As this adjustment is a simple GPU-side operation, it incurs no performance overhead. For the purposes of this image, we opted to remove the intensity multiplier, to show the raw predicted coefficients result of our neural network. Last column indicates the mean squared error (MSE) and 
    standard deviation (STD) between the traditionally evaluated and predicted vertex PRT coefficients.}
    \label{fig:comparison_with_prt}
\end{figure}

Table~\ref{tab:precomp_times} summarizes the five mesh models that were used to 
evaluate our pipeline, along with their respective complexities, expressed in
terms of vertex count. The traditional PRT timing refers to the computation time
needed to generate per-vertex SH coefficients for each mesh. This computation
time is directly influenced by the complexity of the model, as illustrated in
the accompanying table. Specifically, as the number of vertices increases, the
time required for coefficient computation proportionally increases. 
Additionally, the visibility evaluation—which performs occlusion checks for each triangle in multiple directions relative to the rest of the mesh—is computationally intensive. The operation relies on iterative loops, and execution time scales with the number of vertices, leading to a significant performance impact on complex models.

More importantly, as the first 
result (Teapot) suggests, we are able to achieve real-time performance ($<33$msec)
for sufficiently small meshes, i.e., with less than $1000$ vertices, which are 
typical in XR applications. 

On the other hand, the reported Neural-GASh computation time encompasses both 
the transformation of vertex-normal pairs into conformal geometric algebra (CGA) 
multivectors and the subsequent inference of vertex coefficients by the neural 
network, using these multivectors as input representations. Unlike the
traditional PRT approach, which directly computes these coefficients, the Neural-GASh
method leverages pre-trained neural network models to accelerate this process.
The comparison of the visual results of the two methods is presented in
Figure~\ref{fig:comparison_with_prt} and the respective comparison of their
execution times in Table~\ref{tab:precomp_times}.
The last column of Figure~\ref{fig:comparison_with_prt} presents the mean squared error 
and standard deviation of predicted vertex PRT coefficients across five 3D models, 
revealing that prediction accuracy varies with surface complexity rather than vertex count. 
The Statue model exhibits the highest error variability, while the Pawn shows the most 
consistent and accurate predictions.

\begin{figure}[tbp]
    \centering
    \includegraphics[width=1.0\linewidth]{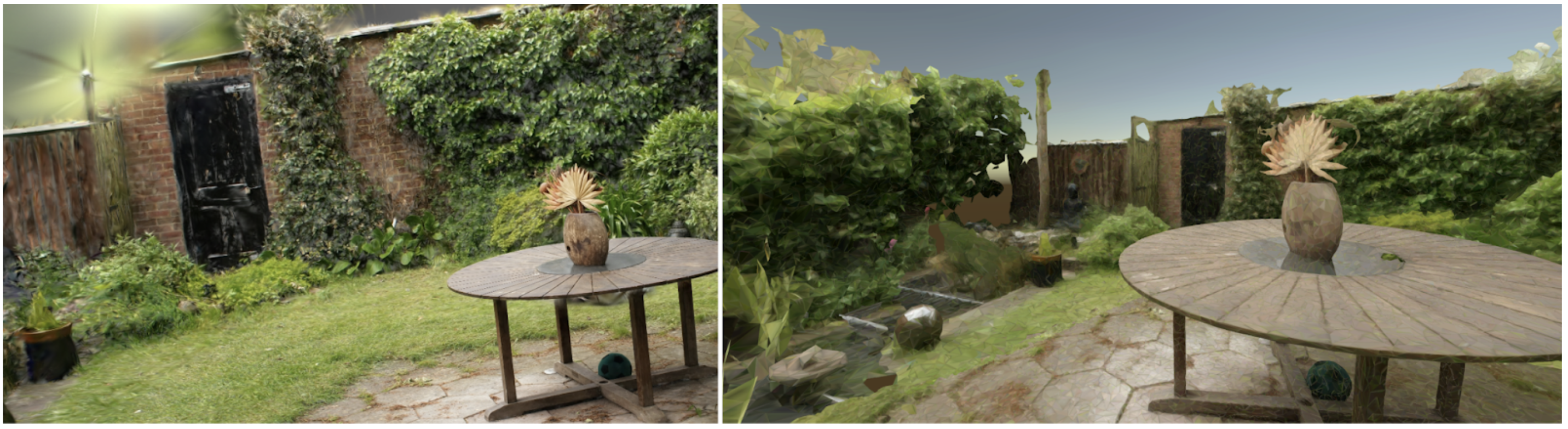}
    \caption{The garden scene from the Mip-NeRF 360 project. On the left is the 
    3D Gaussian Splatting reconstruction, generated from a dataset of 185 images. 
    On the right is the reconstruction using a 3D point cloud, with texture 
    extracted from the Gaussians.}
    \label{fig:gauss_splat_shading}
\end{figure}

\begin{table}[tbp]
    \centering
    \renewcommand{\arraystretch}{1} 
    \setlength{\tabcolsep}{1pt} 
    \begin{tabular}{|c|c|c|c|}
        \hline
        \makecell{\textbf{Mesh} \\ \textbf{Name}} &
        \makecell{\textbf{Mesh} \\ \textbf{Complexity}} &
        \makecell{\textbf{Traditional} \\ \textbf{PRT Time}} &
        \makecell{\textbf{Neural-GASh } \\  \textbf{Time (Unity)}} \\
        \hline
        Teapot   & 672 vertices   & 100.12 s   & \textbf{0.03} s \\
        \hline
        Pawn   & 4892 vertices & 868.42 s & 0.12 s \\
        \hline
        Humanoid & 5340 vertices   & 892.28 s   & 0.20 s \\
        \hline
        Wolf   & 5559 vertices & 962.27 s & 0.15 s \\
        \hline
        Statue   & 49671 vertices  & 10957.75 s  & 1.31 s \\
        \hline
    \end{tabular}
    \caption{Precomputation times for various meshes using Traditional PRT and Neural-GASh. First result (in bold) indicates real-time performance for meshes with less than $1000$ vertices.}
    \label{tab:precomp_times}
\end{table}

\subsection{Gaussian Splats vs Neural-GASh}
The most advanced neural rendering technique to date is the method of 3D
Gaussian Splatting. Although this technique delivers highly realistic results,
a significant limitation remains -- applicability is restricted to static
scenes. As a result the method does not support deformations or animations and
cannot respond to changes in lighting conditions.

For instance, Figure~\ref{fig:gauss_splat_shading} illustrates the well-known garden scene
from the Mip-NeRF 360 publication \cite{barron2022mip}, reconstructed using the
3D Gaussian Splatting technique \cite{kerbl20233d}, which employs Gaussians to
represent the scene geometry.

Since Neural-GASh relies on 3D geometry (vertices and normal vectors), it requires 
the presence and application of such geometric features. For this reason, we 
reconstructed the scene using the SuGaR
technique \cite{guedon2024sugar}, achieving an accurate replica of the scene in
the form of a 3D point cloud, which provides the vertices
and normals required by the algorithm.

\begin{figure}[tbp]
    \centering
    \subfigure[Solitude Night ]{\includegraphics[width=0.235\textwidth]{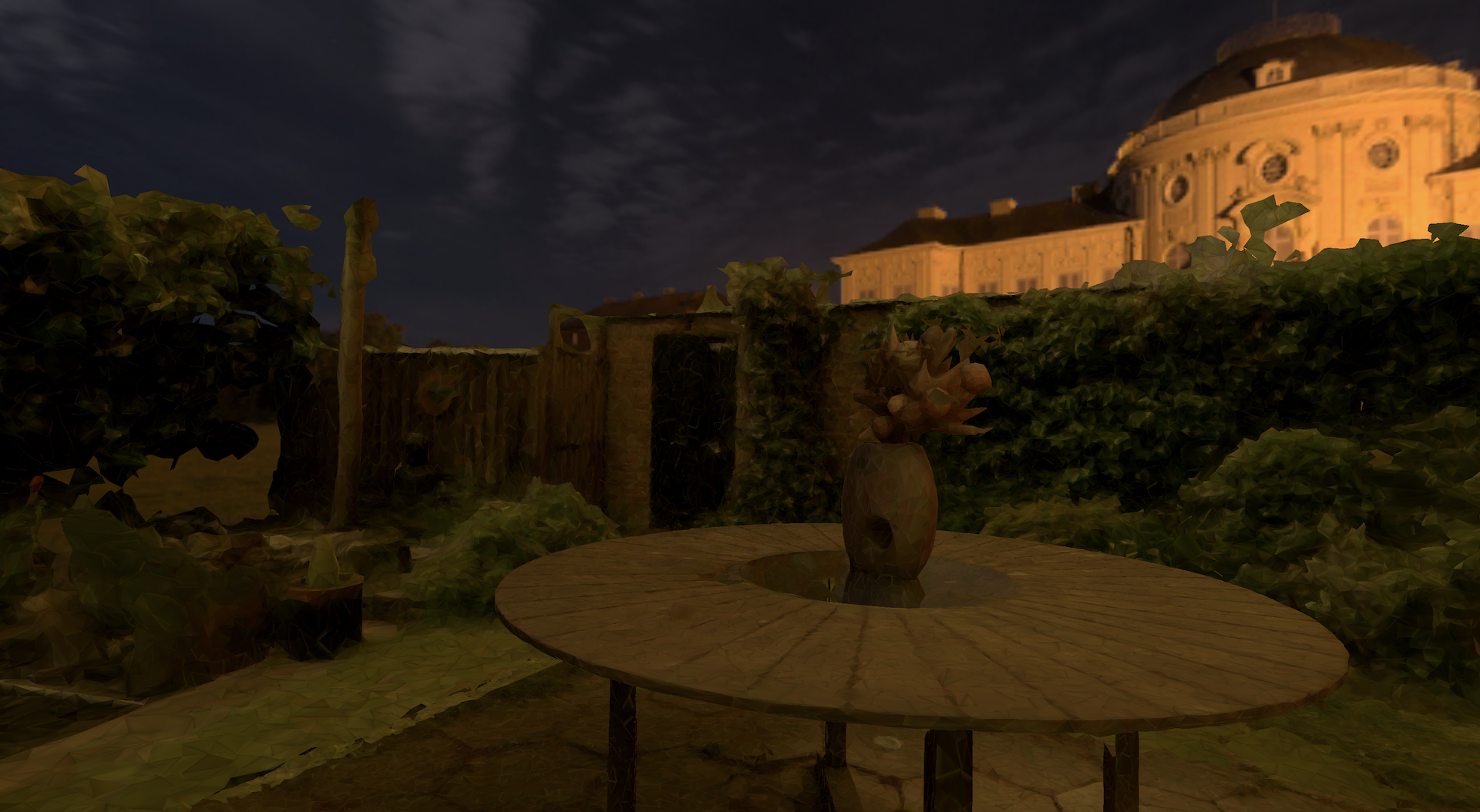}}
    \hfill
    \subfigure[St. Peter Basilica]{\includegraphics[width=0.235\textwidth]{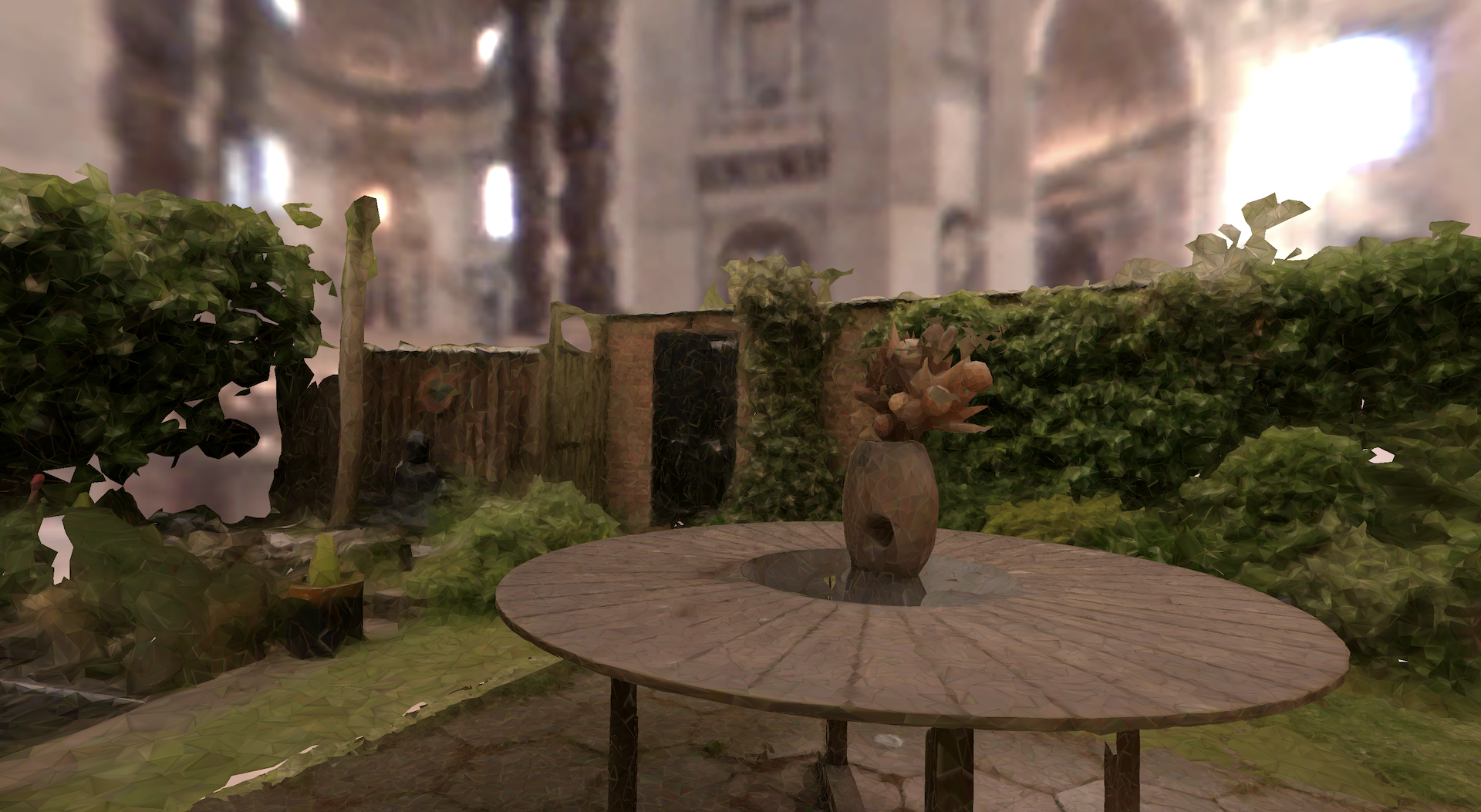}}
    \hfill
    \subfigure[Autumn Field Puresky]{\includegraphics[width=0.235\textwidth]{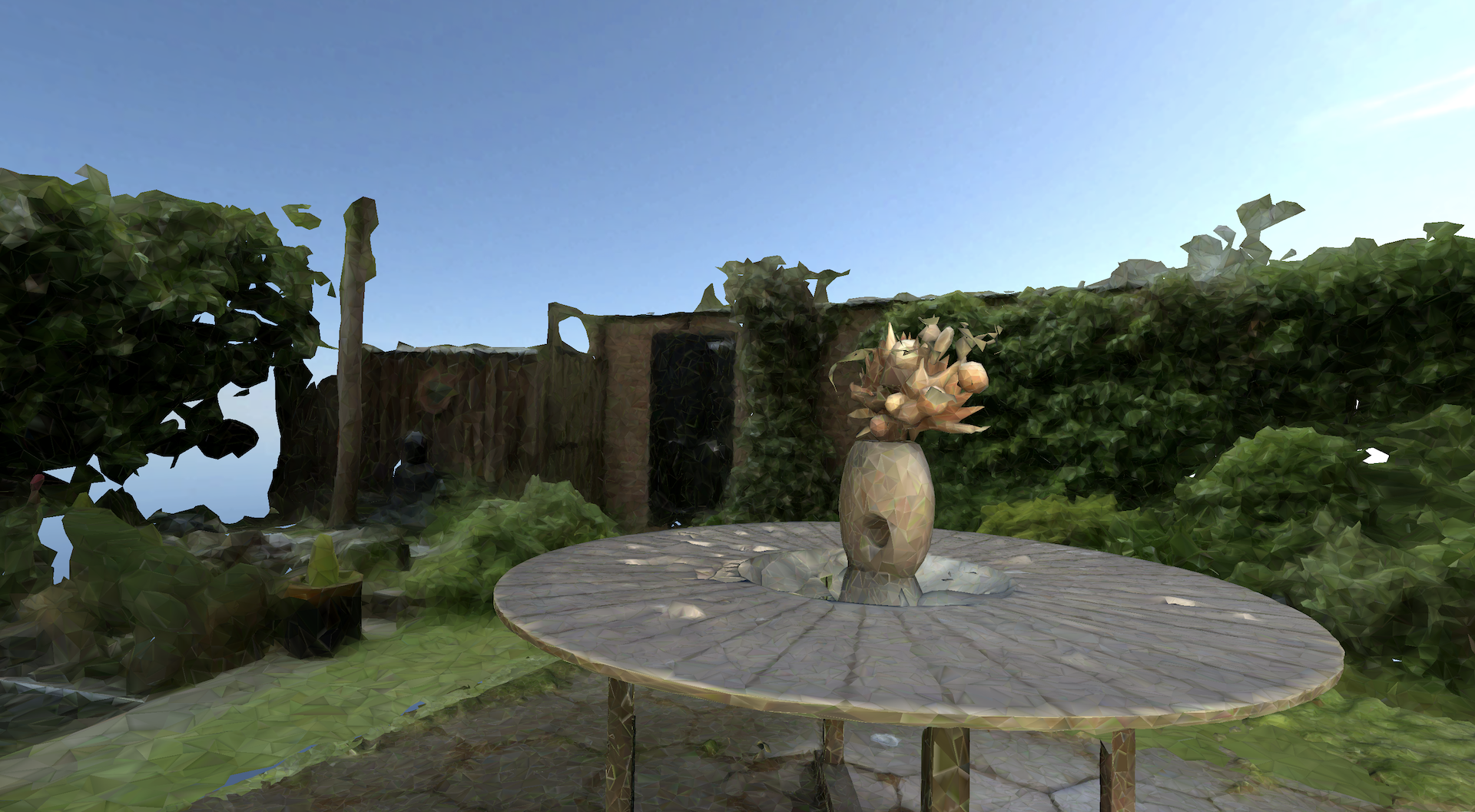}}
    \hfill
    \subfigure[Victoria Sunset]{\includegraphics[width=0.235\textwidth]{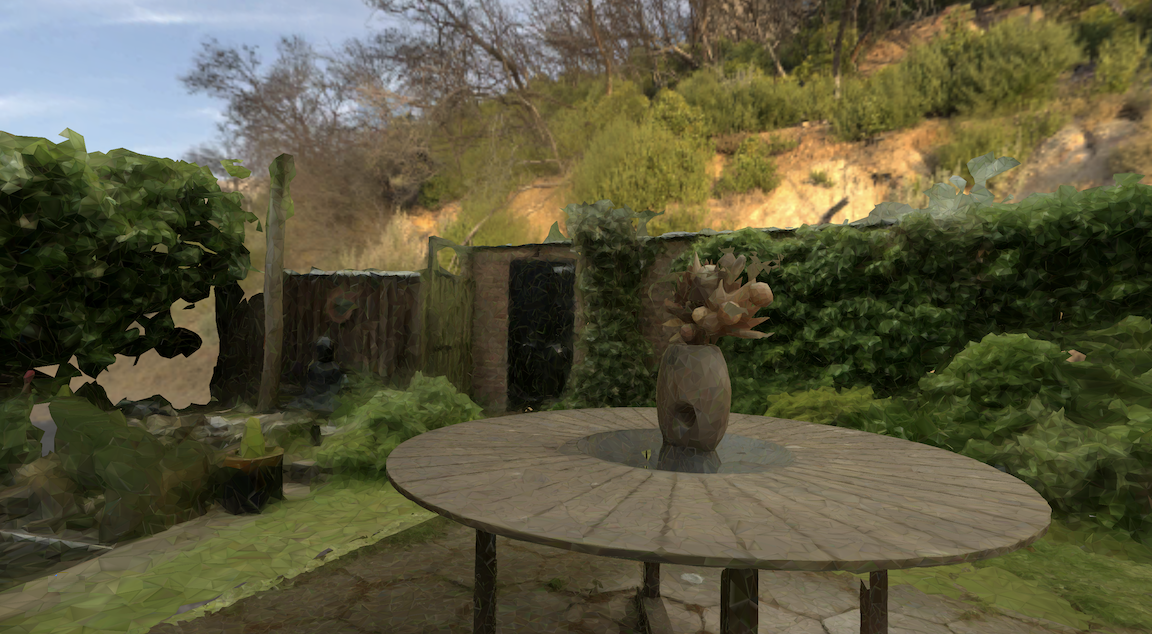}}
    \caption{Lighting results on the reconstructed scene from Mip-NeRF 360, utilizing different radiance maps. The time in seconds needed to switch from one lightmap to another, was between 0.3-0.6 seconds. The lighting changes are performed during program runtime.}
    \label{fig:splats_shade_results}
\end{figure}

\begin{figure}[tbp]
    \centering
    \includegraphics[width=0.235\textwidth]{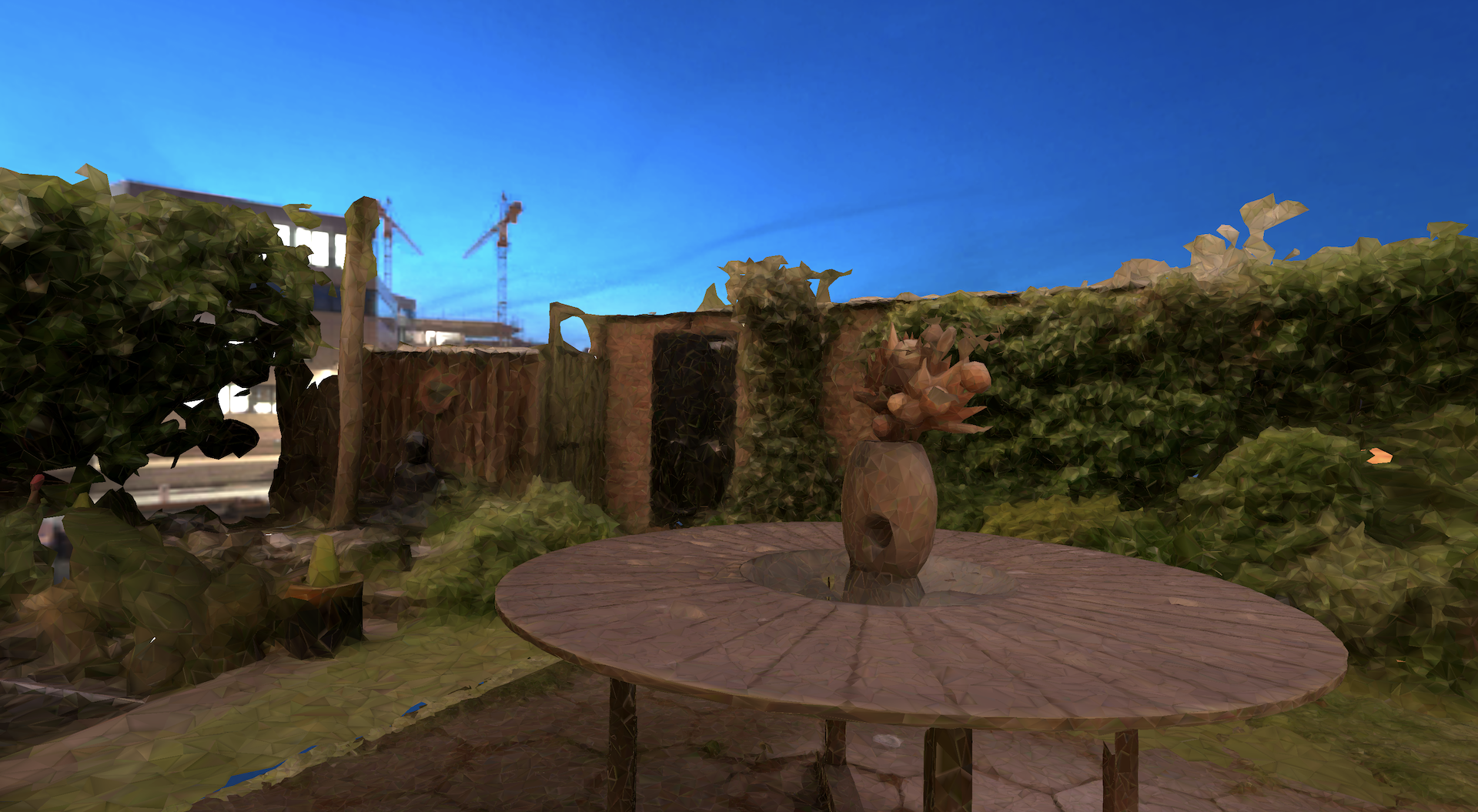}
    \includegraphics[width=0.235\textwidth]{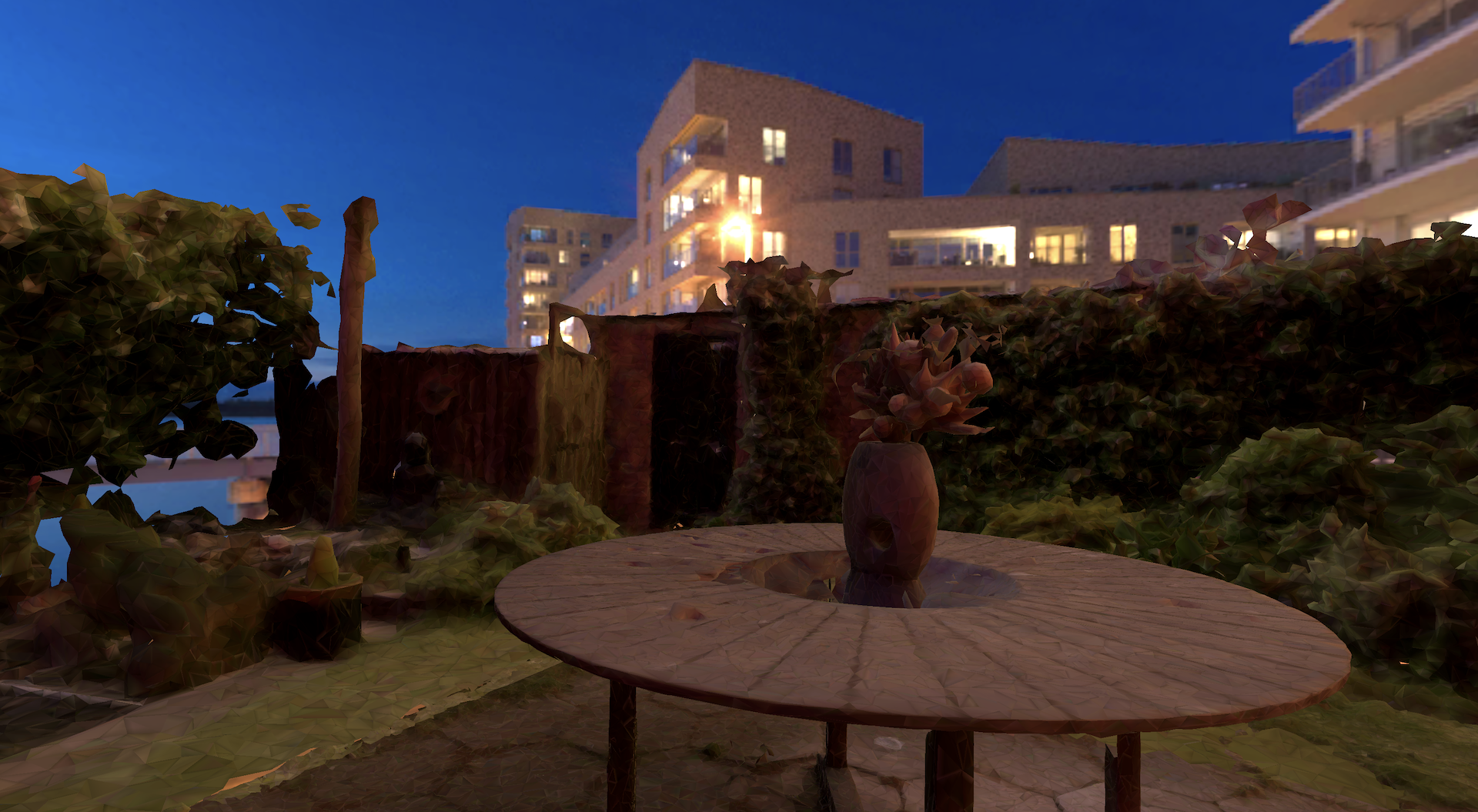}\\ \vspace{2pt}
    \includegraphics[width=0.235\textwidth]{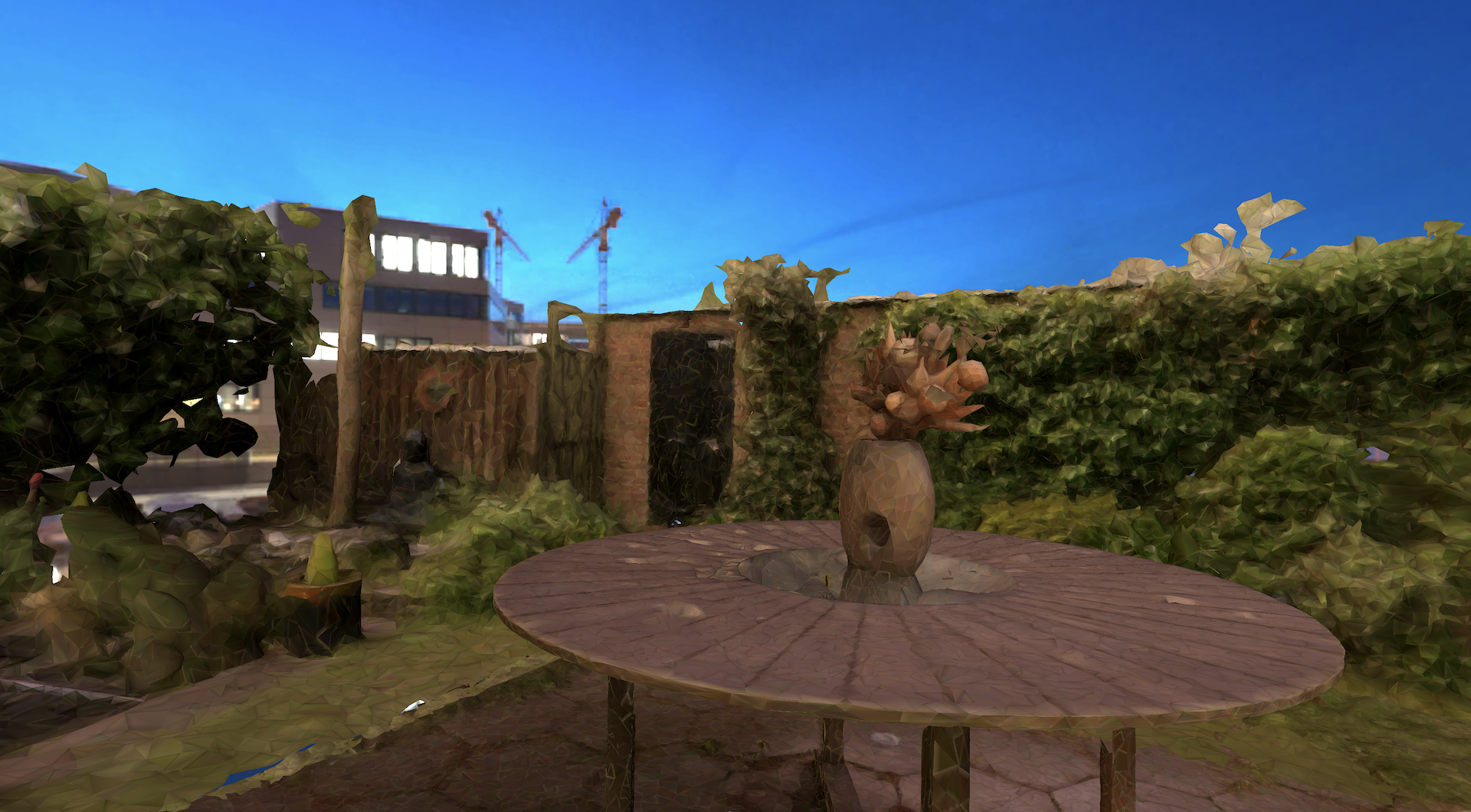}
    \includegraphics[width=0.235\textwidth]{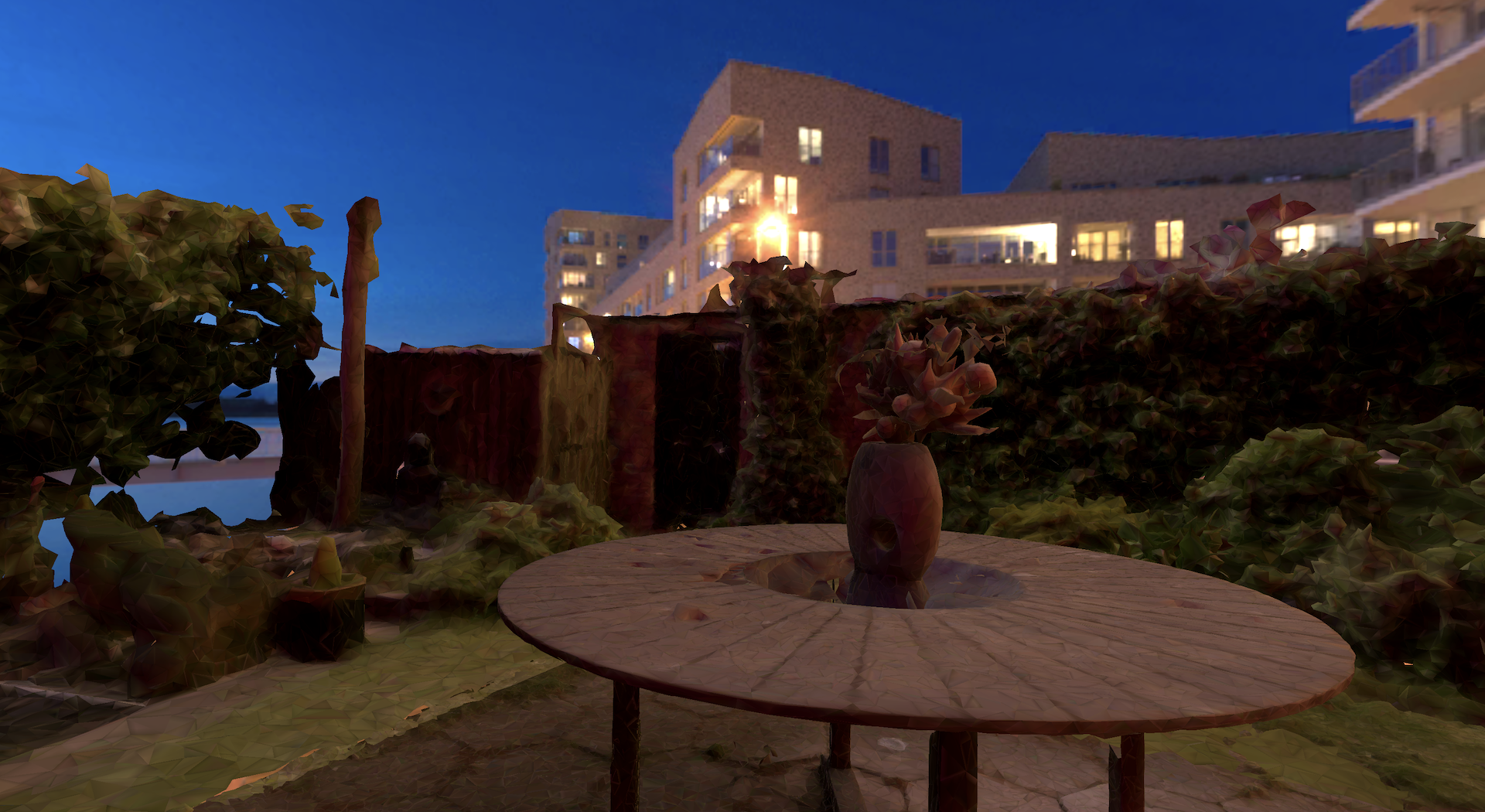}
    \caption{Neural-GASh results for the Mip-NeRF 360 scene, with each instance 
    involving the rotation of light and light coefficients using CGA logic.}
    \label{fig:multiple_gauss_splats}
\end{figure}

As a result, our Neural-GASh system overcomes the limitation of static scenes of 3DGS,
by enabling support for dynamic scenes and real-time lighting control.
This is achieved through the integration of CGA, which facilitates the rotation of SH light
coefficients. This capability is demonstrated in Figure~\ref{fig:multiple_gauss_splats} and
Figure~\ref{fig:splats_shade_results}, where the scene lighting is dynamically 
altered simply by changing the
radiance map in the skybox and rotating it—without requiring a program restart.
Each lighting update is applied in under 1 second on our mid-range evaluation
hardware. To comply with the tensor size limitations imposed by Unity
Barracuda engine, the original geometry was simplified to approximately $80,000$
vertices. This reduced mesh was further partitioned into two sub-meshes of
roughly $40,000$ vertices each, with two parallel instances of the Neural-GASh system
applied. The reported timing reflects the average duration required to update
the lighting (i.e., switch the spherical harmonic light coefficients) across
both sub-meshes. For reference, computing vertex coefficients for the same mesh
using the conventional PRT pipeline required $16,923.55$ seconds.

\begin{table*}[tbp]
\centering
\begin{tabular}{|>{\raggedright\arraybackslash}m{1.8cm}||>{\raggedright\arraybackslash}m{2cm}|>{\raggedright\arraybackslash}m{2.2cm}|>{\centering\arraybackslash}m{2.2cm}|>{\centering\arraybackslash}m{2.2cm}|>{\centering\arraybackslash}m{1.7cm}|>{\centering\arraybackslash}m{2.0cm}|}
\hline
\textbf{Method} & \textbf{Input} & \textbf{Output} & \textbf{Animations/ Deformations Support} & \textbf{Runtime Texture Update Support} & \textbf{Lighting Rotation Support} & \textbf{Game-Engine Compatible} \\
\hline
Traditional PRT \cite{Sloan2002} & 3D Mesh & Colors (float array) & No & Under Conditions & Under Conditions & No \\
\hline
Neural PRT \cite{rainer2022neural} & Pre-Rendered image of mesh & 2D Images (Per Pixel Radiance) & No & No & No & No \\
\hline
DeepPRT\cite{li2019deep} & 3D Mesh & 3D Mesh & Yes & Not Clear & No & No \\
\hline
3DGS \cite{kerbl20233d} & COLMAP Images & Gaussians & No & No & No & No \\
\hline
Neural-GASh & 3D Mesh & Colors (float array) & Yes & Yes & Yes & Yes \\
\hline
\end{tabular}
\caption{Comparison of PRT methods and their capabilities with Neural-GASh.}
\label{tab:method_comparison}
\end{table*}

\subsection{SoA methods vs Neural-GASh}

Table~\ref{tab:method_comparison} presents a comparative analysis of leading
state-of-the-art PRT-based lighting techniques alongside Neural-GASh. For the primary
baseline, we selected the traditional PRT method \cite{Sloan2002}, against which
we benchmark recent advancements in neural rendering approaches, including
Neural PRT \cite{rainer2022neural}, DeepPRT \cite{li2019deep}, and 3D Gaussian
Splatting \cite{kerbl20233d}, the latter representing one of the most prominent
real-time rendering techniques in current use. This comparison focuses on
several key criteria: the nature of the input and output data, support for
animations and deformations, the ability to dynamically update textures at
runtime (in the case of textured models), support for real-time lighting
rotation, and compatibility with game engines for practical deployment in
interactive applications.

Specifically, traditional PRT operates directly on core 3D mesh data, such as
vertex positions and normals. In contrast, Neural PRT processes pre-rendered
images of the mesh and applies PRT-based lighting in image space. Although
DeepPRT utilizes raw 3D mesh input, it extracts a harmonic map from the mesh,
which is subsequently used as a texture for further computation. 3D Gaussian
Splatting reconstructs scenes from COLMAP \cite{schoenberger2016sfm},
\cite{schoenberger2016mvs} images and represents them as Gaussian distributions.
In comparison, Neural-GASh remains closely aligned with the foundational principles of
traditional PRT, operating directly on vertex and normal data without additional
preprocessing, thus maintaining a simple yet effective pipeline.

With regard to the output, traditional PRT produces a color array for the 3D
mesh, encapsulating the lighting information at each vertex. Neural PRT, by
contrast, generates a 2D image in which the mesh is illuminated according to PRT
principles. DeepPRT outputs coefficient images that encode lighting information,
which are subsequently mapped onto the 3D mesh to yield the final appearance. In
the case of 3D Gaussian Splatting, the method produces a set of 3D Gaussians
that collectively represent the reconstructed scene, including its lighting
properties. Consistent with the approach of traditional PRT, Neural-GASh outputs a color
array for the 3D mesh, directly providing per-vertex lighting predictions
derived from the  model.

Animations/deformations are only supported in DeepPRT, since the authors in this
work predict the new coefficients based on the deformations of various meshes.
The rest of the methods 
do not support deformations, since Traditional PRT refers
only to static scenes and objects, Neural PRT works on rendered images and Gaussian 
Splatting utilizes COLMAP dataset (2D images) to reconstruct a realistic but static 3D scene.

In terms of dynamic texture swapping, traditional PRT offers partial support,
contingent upon the specific implementation and the shaders employed; certain
shader configurations may allow real-time texture updates. Neural PRT, operating
entirely on pre-rendered images, inherently lacks support for dynamic texture
modifications. Although DeepPRT might theoretically support this feature,
dynamic texture swapping is not explicitly addressed or demonstrated. 3D
Gaussian Splatting, being a scene reconstruction method based on image inputs,
does not accommodate dynamic texture changes; any modification to textures would
necessitate a complete reprocessing of the scene. By contrast, Neural-GASh employs a
straightforward HLSL vertex-fragment shader that supports texture mapping,
thereby enabling textures to be updated dynamically at runtime. Any change in
the assigned texture is immediately reflected in the rendered output, offering
full support for interactive texture swapping.

Lighting rotation is a critical component of realistic illumination,
particularly for simulating global light sources such as sunlight, which
naturally change orientation over time. To the best of our knowledge, none of
the state-of-the-art methods reviewed here provide support for dynamic lighting
rotation at runtime. Drawing inspiration from the work of
\cite{papaefthymiou2018real}, where lighting rotation was achieved using CGA, we
integrated this approach into Neural-GASh. As a result, Neural-GASh uniquely supports 
real-time
light rotation within a PRT framework, distinguishing itself as the only method
among those compared to offer such functionality.

Moreover, Neural-GASh represents a novel contribution through direct compatibility with game
engines. Specifically, it is the only PRT-based lighting pipeline fully
integrated into Unity, enabling deployment across a wide range of 3D
applications and platforms supported by the engine. This practical applicability
sets Neural-GASh apart, bridging advanced PRT techniques with real-time, interactive
environments.

\section{Limitations \& Future Work}

During our evaluation, Neural-GASh exhibited several limitations affecting
scalability and performance. The current
implementation supports only a single 3D mesh at a time. To process multiple
geometries, a separate Neural-GASh module must be instantiated for each mesh. A
temporary workaround involves merging meshes into a single geometry, but this is
not feasible for animated meshes, where dynamic deformations or skeletal
animations are involved.

Another notable challenge is performance degradation with complex or
high-resolution meshes, particularly when running on resource-constrained
hardware such as mobile devices. Since Neural-GASh performs per-vertex coefficient
prediction followed by lighting computation, the computational cost scales with
mesh complexity, leading to frame rate drops or memory issues under demanding
conditions.

For models with large spatial scale, residual inaccuracies in the
predicted coefficients may still be observed, as illustrated in
Figure~\ref{fig:cats}. While the use of conformal geometric algebra
(CGA) multivectors as input to the neural network significantly
improved prediction accuracy, the results remain suboptimal in certain
cases. This limitation is likely attributable to the training dataset,
which did not include sufficiently large-scale 3D meshes, thereby
limiting the network's ability to generalize across scale variations.

Finally, a challenge of the Neural-GASh pipeline lies in the neural network's reduced
accuracy when predicting coefficients for 3D models that contain large or
complex cavities and regions of pronounced ambient occlusion. 
This limitation stems from Unity Barracuda's constraints, restricting the architecture to static MLPs.
Moreover, Neural-GASh faces Barracuda's tensor size limits, which 
restricts the method to 3D meshes with approximately 100,000 vertices or fewer.
The Python implementation avoids this limitation, handling larger inputs without restrictions.

\textbf{Future Work:} In the future we will focus on optimizing scalability for 
simultaneous multi-mesh processing and lower-end device performance. 
We also aim to expand support for animated
meshes with more than 1000 vertices. While the current
system handles deformations and animations effectively, support for large-scale
animated models (100K vertices and above) is limited by Barracuda’s tensor size
constraints and the existing model architecture. As part of ongoing development, we are 
exploring alternative network architectures to achieve more accurate predictions 
of PRT coefficients, while remaining within the limitations imposed by modern game 
engines. Finally, we intend to augment our dataset with large-scale 3D meshes to enhance the model’s generalization capabilities and mitigate the residual inaccuracies observed in scenarios involving geometries of substantial spatial extent.

\section{Conclusions} In this work, we introduced the Neural-GASh pipeline, a
Unity-compatible system that enables rapid and seamless deployment of the PRT
algorithm to any 3D mesh by eliminating the traditional precomputation step.
This is achieved through the integration of a neural network, which also
facilitates support for mesh animations and deformations. The pipeline's
compatibility with Unity allows for efficient deployment across all platforms
and devices supported by the engine, including mobile VR systems, where performance
remains satisfactory. Integration with CGA further 
enables efficient rotation of light coefficients. Although the method encounters
limitations related to mesh complexity, particularly on lower-end devices, Neural-GASh
establishes a robust foundation for future advancements aimed at optimizing PRT
performance and expanding its applicability in real-time, cross-platform 3D
computer graphics applications.

\textbf{Acknowledgments.} 
This work was partially funded by the Innosuisse Swiss Accelerator (2155012933-OMEN-E), 
and the Horizon Europe Project INDUX-R (GA 101135556).

\bibliographystyle{ACM-Reference-Format} 
\bibliography{Bibliography}      
\end{document}